\documentclass{FrontiersinHarvard}

\usepackage{url,hyperref,lineno,microtype,subcaption}
\usepackage[onehalfspacing]{setspace}

\def\lapprox{\hbox{\lower .8ex\hbox{$\,\buildrel < \over\sim\,$}}}
\def\gapprox{\hbox{\lower .8ex\hbox{$\,\buildrel > \over\sim\,$}}}

\def\keyFont{\fontsize{8}{11}\helveticabold }
\def\firstAuthorLast{{}} 
\def\Authors{Pilar Ruiz--Lapuente$^{1,2}$}

\def\Address{$^{1}$Instituto de F\'{\i}sica Fundamental, Consejo Superior de 
  Investigaciones Cient\'{\i}ficas, c/. Serrano 121, E-28006, Madrid, Spain\\
$^{2}$Institute of Cosmos Sciences, ICC (UB--IEEC),  c/. Mart\'{\i}
i Franqu\'es 1, E--08028, Barcelona, Spain}
\def\corrAuthor{Pilar Ruiz-Lapuente}

\def\corrEmail{pilar@icc.ub.edu}

\begin{document}
\onecolumn
\firstpage{1}

\title[Tycho Brahe's supernova]{On the nature of Tycho Brahe's supernova}

\author[\firstAuthorLast ]{\Authors}
\address{}
\correspondance{}
\extraAuth{}

\maketitle

\begin{abstract}
  At the 450 yr anniversary of its observation,
  the supernova named after Tycho Brahe, 
 SN 1572, can be explained in the terms  
used nowadays to characterize Type Ia supernovae (SNe Ia).
By assembling the records of the observations made in 1572--74
and evaluating their uncertainties, it is possible to
recover the light curve and the color evolution
of this supernova. It is found that, within the SNe Ia family,  
the event should have been a SN Ia with a normal
rate of decline. Concerning the color evolution of SNe Ia, the
most recently recovered records reaffirm previous findings of its being
a normal SN Ia. The abundance studies 
 from X--ray spectroscopy of the whole remnant
  point to a nuclear burning  of the kind of a 
  delayed detonation explosion of a Chandrasekhar--mass white dwarf.
  A tentative single degenerate path to explosion was suggested from
  the exploration of the stars in the field in SN 1572. Though, the
  origin in a double degenerate is being considered as well. 
  Tycho Brahe's supernova, being the first supernova studied by
  astronomers, is still the subject of very intensive debates nowadays. 

\tiny
\keyFont{\section{Keywords:}
  cosmology, stars: supernovae: general, SN 1572, ISM: supernova remnants,
  white dwarfs }
 \end{abstract}

\section{Introduction}

In this review, we will start with the historical data on the 
supernova discovered by
Tycho Brahe, SN 1572.  We give an account of  its 
 light curve and cosmological charaterization. 
 We discuss as well the explosion mechanism of the supernova and
 the binary path leading to the
the explosion, taking into account the most recent 
work.  This 
article aims to address what
is known about the nature of Tycho's SN. Though, due to
constraints on the length, not all the contributions
in relation to this topic
can be discussed. We will try to present the most recent overall view.

\noindent
   We will begin with an introduction of SNe Ia as calibrated candles.
  The understanding of SNe Ia as empirical tools in cosmology had an enormous
 boost at the end of the past century.  Pskovskii (1977, 1984) first suggested 
 a correlation between absolute magnitude at maximum 
and rate of decline of the
 light curve. The faster the decline of the light curve the dimmer 
   the SN Ia  and the slower the decline  the brighter the SN Ia.
  The follow--up of SNe Ia  with modern digital CCD detectors, as done 
  in the Calan-Tololo search, 
  confirmed that such correlation had a low intrinsic dispersion.
  Extensive and accurate observations collected by Phillips, Hamuy 
  and collaborators allowed to build up the mathematical
  expression. Phillips (1993) first quantified it using a small number
  of SNe Ia. After that,
  Hamuy et al (1996a,b) enlarged the study and obtained a more significant 
  fit.     
The relation has the form: 

\begin{equation}
M_{MAX} = a + b[\Delta m_{15}(b) - 1.1]
\end{equation}

\noindent
where M$_{MAX}$ is the absolute magnitude at maximum,  a and b are constants,
 and $\Delta m_{15}$ is the number of magnitudes 
increased in 15 days after maximum. As the first results were given by 
Phillips (1993), this is called the Phillips relationship.

\begin{figure}
\centering
\includegraphics[width=0.6\columnwidth]{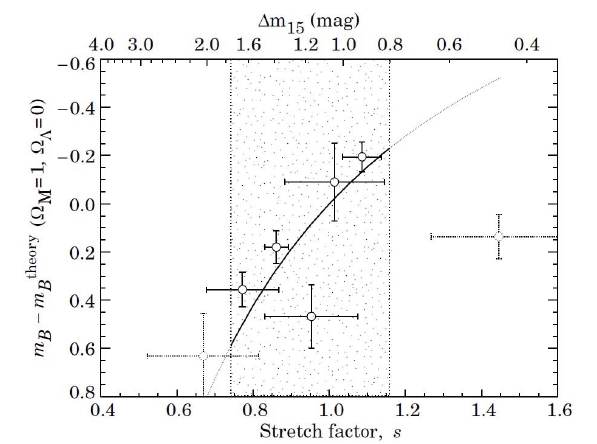}
\caption{ 
m$_{B}$ magnitudes versus the best-fit stretch factor, $s$,
for high-redshift supernovae.
The upper axis gives the equivalent values of $\Delta m_{15}$
$=$ $1.96(s^{-1}-1) + 1.07$. Figure from Perlmutter et al.
(Supernova Cosmology Project) in {\it Thermonuclear Supernovae}
(ed. P. Ruiz--Lapuente  et al. 1997). $\copyright$
Kluwer Academic Publishers. Reproduced with permission.
}
\label{fig:1}
\end{figure}

\noindent
The {\it Supernova Cosmology Project} used for that maximum brightness--
rate of decline of the light curve the so-called
  {\it stretch factor}, $s$ (Perlmutter et al. 1997, Goldhaber et al.
 2001), measuring the amount of 
broadening or narrowing of the light curve up to 
60 days after maximum in relation to a template light curve (see Figure 1
from comparison of $\Delta m_{15}$ and $s$).
In this way, they were able to use the corrected to template 
light curves for cosmology, $m^{\rm eff}_{B}$:

\begin{equation}
  m^{\rm eff}_{B} \equiv  m_{B} + \alpha(s - 1)
\end{equation}

\noindent
where 
$\alpha$ is the best--fit value of the stretch-luminosity slope from the fit to
the primary low-extinction subset (Knop et al. 2003)\footnote{This is applied
after correcting from extinction as implied by the reddening of the SN Ia.
In a more recent version, color changes due to reddening or intrinsic
color of SNe are treated equally in the SALT2 standarization. 
The coefficients for stretch, x1, and color, c, go respectively in a 
  $\alpha \times x_{1} - \beta \times c$ correction to the observed magnitude.
(Some more recent parametrizations add additional terms in the equation). 

\noindent
There are relations between x1, the stretch s from the {\it Supernova
  Cosmology Project} and the $\Delta m_{15}$ : 

\begin{equation}
  s(SCP) = 1.07 + 0.069 x_{1} - 0.015 x_{1}^{2} + 0.00067 x_{1}^{3}
\end{equation}

\begin{equation}
\Delta m_{15} = 1.09 - 0.161 x_{1} + 0.013 x_{1}^{2} - 0.00130 x_1{1}^{3}
  \end{equation}
 }.

\noindent
On the other hand,  the {\it High-z 
Supernova Team} employed the MLCS ({\it Multicolor Light Curve Shapes}: Riess 
et al. 1995, 1996) method.
   Riess, Press \& Kirshner (1995) give an alternative 
  way to express the intrinsic luminosity of Type Ia supernovae
  as related to the rate of decline of the light curve. They fit
  with an overall shape parameter the evolution in luminosity
  before maximum to well past maximum. 
   Their parameterization, and the one used by Hamuy and
  collaborators (1996a;b) give a comparable scale of magnitudes
  for specific SNe Ia The most recent version by this method is the MLCSk2 (Jha, Riess \& Kirshner 2007).

\noindent  
 Several SNe Ia had been discovered 
after maximum, so the peak brightness had to be extrapolated by comparison with 
light-curve templates. These SNe were at distances corresponding to
redshifts $0.01 < z < 0.1$. During the 90's,  the {\it Supernova Cosmology 
Project} and the {\it High-z Supernova Team}, had started to observe SNe Ia at 
higher redshifts, $0.18 \le z \le 0.83$, to measure the evolution of the 
Hubble parameter, $H(z)$. 
From two different samples of high-redshift SNeIa and 
using these different methods of light-curve fitting, the two groups reached
the same conclusion: the expansion of the universe is accelerating (Riess et 
al 1998; Perlmutter et al. 1999).

\noindent
Tycho SN 1572 looked like a normal SN Ia, i.e, did not show to be 
neither overluminous nor underluminous if the historical records of
the SN were fit with the {\it stretch} relationship for SNe Ia 
(Ruiz-Lapuente 2004, hereafter called R04).                   
The light curve shape has been confirmed and the color evolution from a
typical SN Ia has been proved in more recent historical studies
(Neuhaeuser 2022).

\noindent
In recent years, new types of explosions akin to 
SNe Ia have been 
observed, which do not belong to the ``normal'' class, they being either 
overluminous or underluminous and thus not following the Phillips relationship.

\noindent
The historical view of Tycho's SN has been expanded with historical 
research that will be mentioned in Section 2. 
The light curve, color and  extinction of the supernova are treated in
Section 3.
Concerning the search for the progenitor and the explosion mechanism,
 there have been recent revisions and the current state of the question will
 be addressed in Section 4.

\section{Historical records}

\noindent
SN 1572 was well observed in Europe (as well as in the 
Middle--East and Far--East) 
for almost two years.   
It added a new aspect to the debate at the time over 
the Aristotelian cosmological views, as it forced to reconsider the
immutability of the heavens and the solid nature of the
celestial spheres: the ``star'' gained
brightness and lost it during a period of two years, but it showed  
no detectable parallax. 
According to prevalent views about the heavens,
mutability would only happen in the sublunar region. This was even the
place where comets were assumed to originate. The appearence of SN 1572
challenged the order of the celestial spheres. The observers
who followed it up\footnote{The observers who mostly contributed to measure
the position and luminosity,
 Tycho Brahe, Thomas Digges, Thaddeus Hagecius, 
Michael M\"astlin, Jer\'onimo Mu\~noz, 
Caspar Peucer \& Johannes Pr\"atorius held very different views on
the meaning of SN 1572.  
A comparison of their measurements and an account of
 their views is given elsewhere.  Recently Neuhaeuser (2022) has provided
more data from German, Italian and Czech astronomers of the epoch such as 
 Adam Ursinus 
(also called Adam B\"ar), Francesco Maurolyco, Cyprian Leowitz and Georg Busch.}
took sides with respect to established Aristotelian
views (Ruiz--Lapuente 2005). Today we can
still use their observations to see whether that supernova would be
of use for cosmology in a different way, as a distance indicator if
seen by observers billions of lightyears away from us. It is indeed 
possible to have a clear idea of where SN 1572 stands among its class.

\noindent
After several centuries of questioning the nature of SN 1572, 
the identification as a Type I supernova came through
the revision of the light curve done by Baade (1945) and based on  
data taken or quoted by Tycho Brahe (1603a). 
Before that time, there were still speculations on whether it was a
variable star of some kind, a nova (Morgan 1945) 
and  the suggestion of 
its cometary nature was still considered (Lynn 1883). 
The cometary idea expressed in 1573 by 
Jer\'onimo Mu\~noz, was based on the fact that the event is 
aligned with the Milky Way, so 
its decay in luminosity could be explained if it were a
comet born among the stars that would first  
approach and then move away from us just along the line of sight. 
 Tycho's {\it stella nova}
 lies indeed only  49--98 pc above the Galactic plane.

\noindent
In modern times, its comparison with other SNe allowed its classification
as a Type I supernova (Baade 1945). Later on, this class was shown to 
contain events of very different nature: those identified
with the explosion of white dwarfs (Type Ia, SN Ia)
and those corresponding to the
collapse of massive stars whose envelope had lost its  
hydrogen content in the interaction with a binary companion (Type Ib or c,
SN Ib or SN Ic, respectively).  
SN 1572 was of the Type Ia class as discussed by de Vaucouleurs
(1985) and by van den Bergh (1993). No further doubts that it is 
a SN Ia can now 
be held in view of the X--ray spectrum of its remnant,
 which clearly differentiates
SNe Ia events from SNe coming from the collapse of massive stars 
(Hughes et al. 1995). A first comparison of its luminosity at maximum
with the bulk of SNe Ia 
would have led to think that it was fainter than normal SNe Ia. 
van den Bergh (1993) considers whether it could be a peculiar, 
subluminous SN Ia, like SN 1991bg. However, SN 1572 was heavily obscured.
It was reddened by 
E(B--V)=0.6$\pm$ 0.04 as it corresponds to the reddening of the stars 
near its position (R04).
After taking into account 
the extinction undergone across the Milky Way, as well as the
measurement of its decline rate, it is concluded that SN 1572 was not a 
91bg--like event. Neither was it 
an overluminous SN Ia like SN 1991T or similars, but
rather an event in the middle of the SN Ia class (R04).

\section{Evidence on visual, color evolution and late decline}

\noindent
The SN 1572 data are compared in  R04
to templates using the stretch factor {\it s}
for the characterization of the rate of decline (Perlmutter et al. 1999; 
Goldhaber et al 2001; Nobili et al. 2003). 
As already mentioned the stretch factor {\it s} method 
was introduced by the {\it Supernova Cosmology
Project} to quantify the decline
rate of the supernova from data extending up to 60 days after maximum. Even 
in absence of a measurement of the brightness at maximum, the method
produces a fine description of the supernova within the 
family of decline rates. It makes sense to use the {\it stretch}
original characterization, as we know what is the reddening suffered
by the SN Ia. When we compare magnitude data points and
magnitude limits by this method, we clearly see that SN 1572 was not a
fast decliner. The best fit corresponds to a stretch factor $s \sim$ 0.9.
 SN 1572 was an event, for instance, very
 similar to SN 1996X,
with $s$ = 0.889. A comparison is made in Figure 2, with SN 1996X and with
the subluminous SN 1991bg ($s$ = 0.62). The SN 1572 points are shown together
with a template light curve with $s$ = 0.9. We see that there is a clear
discrepancy with the fast-declining subluminous SN 1991bg. A comparison with
the broad light curve of the overluminous SN 1991T shows a
similar discrepancy, of the opposite sign.

\begin{figure}
\centering
\includegraphics[width=0.6\columnwidth]{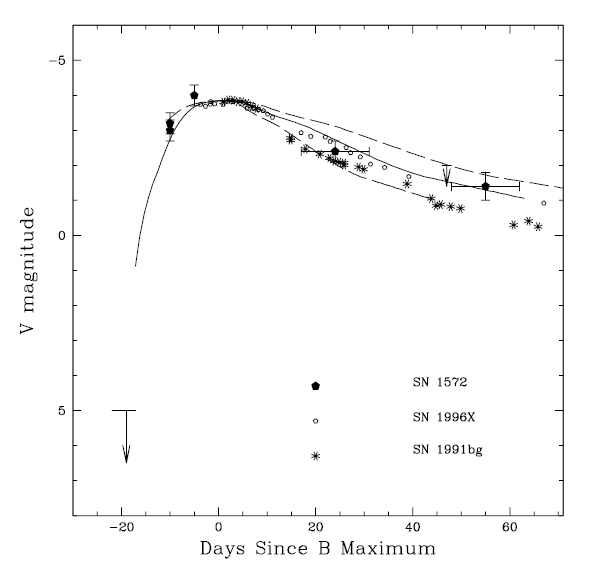}
\caption{The visual light curve of SN 1572 till 60 days. The solid curve is the V light
curve of a s = 0.9 SN Ia, which gives the best account for the decline. Such stretch factor
is typical of normal SNe Ia. We show for comparison the V light curve of the normal SN Ia
SN 1996X, whose stretch factor is s = 0.889 and of the fast–declining SN 1991bg. SN 1572
was significantly slower than SN 1991bg. The light curves plotted in dashed lines are the
templates of 91bg–type events and 91T–like events, which depart significantly from the data. Figure in R04. $\copyright$ AAS. Reproduced wih permission. 
}
\label{fig:2}
\end{figure}

\begin{figure}
\centering
\includegraphics[width=0.8\columnwidth]{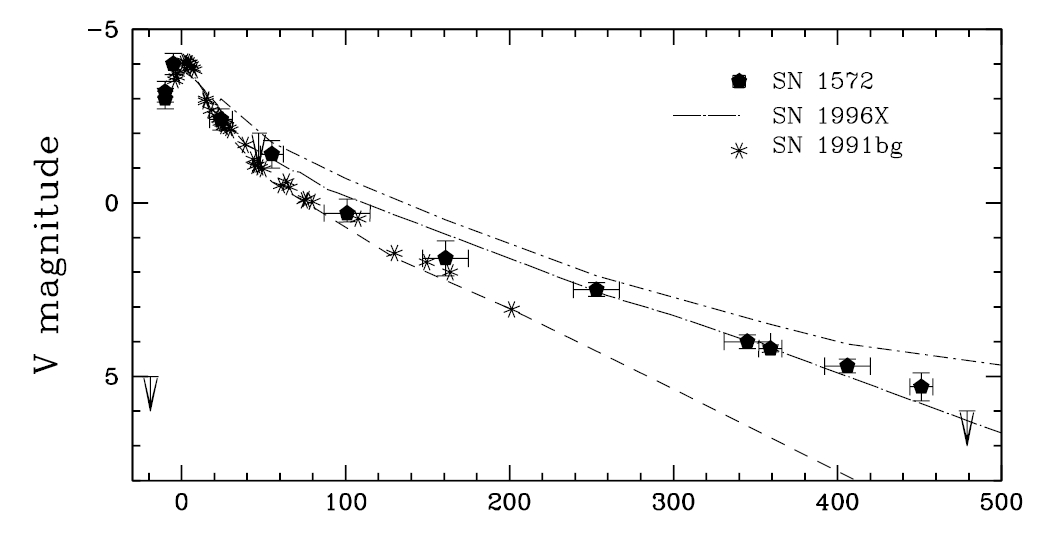}
\caption{The visual light curve of SN 1572 till 500 days. 
Its late rate of decline is the one of normal SNe Ia. It is
very similar to the decline of the s $=$ 0.889 SN 1996X. The 
visual data of SN1991bg and the template light curves of this SN Ia
and SN 1991T are shown for comparison. Figure in R04. $\copyright$ AAS.
Reproduced with permission. 
}
\label{fig:3}
\end{figure}

\smallskip
\noindent
The late decline of SN 1572, from 100 to 450 days after maximum, which is slow
and similar to that of the bulk of SNe Ia, is an additional proof  that it
was a normal one. SN 1572 declined by 1.4 mag in 100 days, similar to
SN 1990N and other normal SNe Ia, with decline rates of 1.38-1.5 mag, also in
100 days. In contrast, SN 1991bg declined again faster over the same period
(see Filippenko et al. 1992; Leibundgut et al, 1993; Ruiz-Lapuente et al. 1993;
Turatto et al. 1996). SN 1572 is clearly not a peculiar ''91bg''--like SN Ia. 

\noindent
Now using the whole light curve, we consider the templates for a normal event
with $s$ = 0.9 (like SN 1996X), for a subluminous SN 1991bg type and for
an overluminous event like SN 1991T, all of them matching the respective
available observations. Up to 60 days after maximum, such templates coincide
with those of Hamuy et al. (1996b) and with the SCP templates for the same
$s$. At later times, they follow the available late-time photometry
(Schmidt et al. 1993; Salvo et al. 2011).

\smallskip
\noindent
As reported in  R04, the fit of the
SN 1572 data to the $s$ = 0.9 template has an acceptable
$\chi^{2}$ of 14.44 for 10 degrees of freedom, whereas the fit to the
$s$ = 0.62 template of a fast-declining, subluminous SN Ia (SN 1991bg)
has a $\chi^{2}$ of 53.55 for 3 degrees of freedom (the two premaximum points
and the last four points having been omitted due to the lack of data for
such stages, in those subluminous events), which is exceedingly high.
At the opposite end, the fit to the $s$ = 1.2 template, for slow-declining,
overluminous SN Ia like SN 1991T, has a $\chi^{2}$ of 82.99 for 10 degrees of
freedom, which is also too high.

\smallskip
\noindent
We show in Figure 3 the total visual light curve of SN 1572, extending to
almost 500 days after maximum light, together with the SN 1991bg and
SN 1991T templates. The supernova had not yet leveled-off at 480 days past
maximum, according to the last upper limit given by Tycho Brahe. It probably 
leveled-off later, at around 500 days, since there is a
light echo of the SN  produced by dust clouds nearby, discovered by Krause
et al. (2008). 

\noindent
The production of
echos  depends on the distribution of the clouds, both with
respect to the SN and to the observer. So, for instance, in SN 1986G, though
 being heavily reddened, no echo was observed (Schmidt et al. 1993). 
SN 1991T and SN 1998bu have shown a slowing down in the V magnitude rate of
decline at some 400 days after maxmum, that being due to light echos
(Schmidt et al. 1993; Cappellaro et al. 2001). The leveling-off took place
at 500 days for those SNe Ia, at some 10 mag from maximum. That, 
in the case of SN 1572, similar to SN 1998bu, 
would place the leveling at V = 6 mag, approximately at the limit for the
naked eye.  SN 1998bu is, in fact, a SN Ia very similar to SN 1996X and SN 1572,
its reddening being $E(B-V) = 0.32 \pm 0.04$ mag only (Hernandez et al.
2000; Cappellaro et al. 2001), which is half the reddening of
SN 1572. One could say that SN 1998bu is a Tycho Brahe with half the extinction
suffered by the historical SN 1572. (See next subsection about the similarity of
SN 1572 and SN 1996X and SN 1998bu as shown in the spectrum observed in the echo). 

\smallskip
\noindent
In addition, the late light curve of SN 1572 tells us of the energy deposition
due to the decay of $^{56}{\rm Co}$ to $^{56}{\rm Fe}$ in the SN ejecta.
The decline being slow points to the explosion of a Chandrasekhar-mass white
dwarf and the late decline in $V$ is indicative of a significant deposition
of energy by the positrons produced in the decay
$^{56}{\rm Co} \rightarrow\ ^{56}{\rm Fe} + e^{+}$. Indeed, from 200  days
after the
explosion, the positrons become the main luminosity source. Departure from
full positron trapping in the ejecta is always observed, the reason
being incomplete confinement of the positrons and/or incomplete thermalization
of their energies.
Several causes can produce diversity in the late-time SN light curves:
differences in the nucleosynthesis yields and the kinematics of the ejecta,
degree of mixing and configuration and intensity of the magnetic field
(see Ruiz-Lapuente \& Spruit (1998). 
It is found that departures of the order of 10-15\% from full trapping of the
positrons at 400 days time can be explained by the distribution of radioactive
material, but larger departures, of the order of 30-40\% or more, require a
lack of confinement of the positrons by the magnetic field or even an
enhancement of their escape due to a radially combed magnetic field. In the
particular case of SN 1991bg, the very fast drop of the light curve indicates
not only an ejected mass below the Chandrasekhar mass but also the absence of
a tangled magnetic field.

\noindent
Milne et al. (1999) made a comparison of the predictions from several
explosion models for SNe Ia with a sizeable sample of bolometric light curves.
No bolometric light curve for SN 1572 exists, the data being only for the
$V$ band, but we can draw analogies with SNe Ia for which we know the bolometric
data.

\smallskip
\noindent
As said above, the late decline of SN 1572 is an important proof, confirming
that Tycho's SN falls in no way within the estimated 16\% (Li et al. 2001) of
intrinsically subluminous SNe Ia. Such low-luminosity class likely arises
from a peculiar type of SNe Ia explosions, likely
 ejecting a smaller amount of mass than
normal SNe Ia (Ruiz-Lapuente et al. 1993).

\subsection{Light echo of SNe Ia}

\noindent
 Sometimes it is possible to know details about
 a SN Ia long after its light
 has gone away (see Figure 3), through the study of the echo coming from dusty
 regions around 
 the supernova. The light traveling to a dusty region and beeing scattered and 
 absorbed and re-emitted can take some hundred years to arrive. 
 This is the case for Tycho's SN.

 \begin{figure}[h!]
\centering
\includegraphics[width=0.7\columnwidth]{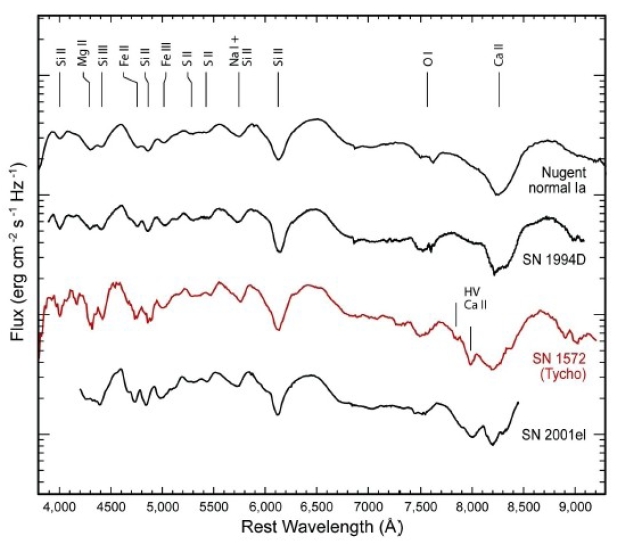}
\caption{Spectrum of the light echo of SN 1572 compared with other
normal SNe Ia.  The spectra are weighted averages around maximum light.
Figure from Krause et al (2008). Courtesy of T. Usuda. $\copyright$ Springer
Nature. Reproduced with Permission.}
\label{fig:4}
 \end{figure}
 
\noindent
Through the study by Krause et al (2008), we
  know that the supernova was a normal one
  (Krause et al. 2008; Usuda et al. 2013). The spectrum taken of the echo
reveals the ion absorptions in full detail and
 it fits perfectly with a normal SN Ia. 
Moreover, that paper establishes that the spectrum of SN 1572 
matches the comparison spectra of four well observed
normal SNe Ia (1994D, 1996X, 1998bu, 2005cf) (Krause et al. 2008). That
coincides with the conclusions derived from the light 
curve (R04).  

\noindent
In the same way that the light curves of superluminous SNe Ia and subluminous
SNe Ia differ, the first ones being brighter and slower in their
rise and decline while 
the second ones being less luminous and faster in decline, they also differ 
in spectra. Superluminous SNe Ia lack a well defined Si II $\lambda$ 6355 
 \AA \ absorption feature at maximum light,  though the subsequent evolution is 
 similar to normal SNe Ia.  On the other hand, subluminous SNe Ia
 show a characteristic absorption 
near 4200 \AA \ atributed to Ti II, near maximum light.

\noindent
The spectrum coming from the echo represents those of added epochs corresponding
to extent of the scattering cloud. So, typically one compares the echo spectrum
with the spectra of a SN Ia time averaged over the brightness peak of the 
light curve. The dominating characteristics are those of the brightness peak, 
though it is reasonable to make the weighting by 
brightnesses of the spectra
from 0 to 90 days after explosion, as it is 
 done for several SNe Ia spectra derived
from echoes.

\subsection{Reddening and color}

\smallskip
\noindent
The most direct estimate of the interstellar reddening in the direction of
Tycho's SNR is provided by the measurement of the reddening and extinction
of the stars close to the centroid of the remnant and at similar distances.
The distance to Tycho SN is in the range of 2 to 4 kpc (de Vaucouleurs (1985)
reviewed values obtained by various methods and they lie in this range).
 An estimate from Tian anf Leahy (2011) based on the modeling of the kinematics of the region  gives
2.5-3.0 kpc. The stars near the centroid have
average reddenings of $E(B-V) = 0.6 \pm$ 0.04 mag. The quoted value comes from
the program to search the companion stars of Galactic SNeIa (Ruiz-Lapuente et
al. 2004, hereafer RL04). The candidate stars for SN 1572
are within an angular distance from the
centroid of the remnant including the uncertainty on the site of the
explosion plus the shift in position corresponding to traveling perpendicularly
to the line of sight for 431 years, at the velocities expected for the fastest
moving companions.

\smallskip
\noindent
The stars were modeled to obtain the stellar atmosphere parameters $T_{\rm eff}$,
log $g$ and  [Fe/H], plus the distance and $E(B-V)$. Radial velocities were
measured from the spectra. The program stars cover 35\% of the radius of
the remnant, and their reddenings span from $E(B-V)$ = 0.50 mag to $E(B-V)$ =
0.8 mag, the values increasing with the distance.

\smallskip
\noindent
The mean reddening above gives an extinction $A_{V}$ = 1.86 $\pm$ 0.12 mag,
adopting a $R_{V}$ of 3.1 (Sneden et al. 1978; Ricke \& Lebofsky 1985).
The Galactic extinction data from COBE/DIRBE (Drimmel \& Spergel 2001) give
$A_{V}$ = 1.77 mag in that direction, the maximum Galactic extinction there
being $A_{V}$ = 1.90 mag. Therefore, the extinction measured at the distance of
SN 1572 does agree with the COBE/DIRBE values. Once the apparent brightness is
corrected for extinction, it confirms Tycho's SN
as a normal SN Ia and not a subluminous one.

\smallskip
\noindent
The historical records on the color evolution of SN 1572 can equally be
corrected for reddening and obtain the intrinsic color evolution.  Two
months after maximum light, all SNeIa show a similar evolution in color, and
there is a well established law, with low intrinsic dispersion, valid for the
period froom 30 to 90 days, studied by Phillips et al. (1999) (Ph99) and based
on the work of Lira (1995):

\begin{equation}
(B-V)_{0} = 0.725(\pm0.05) - 0.0118(t_{V} - 60)
\end{equation}

\noindent
where $t_{V}$ is the time from visual maximum.

\smallskip
\noindent
Two months after discovery, the color of SN 1572 was reported to be similar to
that of Mars and Aldebaran, which means that $B-V$ was in the 1.36-1.54 mag
range. Other color estimates, previous and subsequent to this one, are given in
Table 2 of R04. New historical records have been recently
added by Neuhaeuser (2022). Based on the observations of Adam Ursinus, Georg
Busch and Cyprian Leowitz, this author asserts that by the end of November the
supernova appeared withish or silver, becoming yellowish at the end. These
new three points
 confirm that the SN could not be of the SN 1991bg type, since it should
have been reddish in that case. Instead, the points fall where they are
expected for a normal SN Ia.

\smallskip
\noindent
After correction of the observed colors for our measured reddening of
$E(B-V)$ = 0.6 $\pm$ 0.04 mag, the intrinsic color at 55 $\pm$ 10 days becomes
$(B-V)_{0}$ = 0.76 $\pm$ 0.24 mag, in very good agreement with the expected
$(B-V)_{0}$ = 0.78 $\pm$ 0.15 mag for the given epoch, thus fitting very well
in the Ph99 law above. Nobili et al. (2002) have shown that the color
evolution of the bulk of the SNe Ia does follow the law, with only a low
dispersion of 0.1 mag in the tail.

\smallskip
\noindent
Before maximum, the corrected color of SN 1572 would have been $(B-V)_{0}$ =
0.22 $\pm$ 0.29 mag, which is consistent with the fact that normal SNe Ia
have $(B-V)_{0}$ $\sim$ 0 mag. In contrast, SN 1991bg, as well as other
subluminous SNe Ia, clearly deviate from the standard color evolution at
early epochs also, they being intrinsically redder at maximum, with
$(B-V)_{0}$ = 0.6 mag (Leibundgut et al. 1993; Ph99). Additionally,
the color evolution for a $s$ = 0.9 SN Ia agrees well with the SN 1572 data.

\smallskip
\noindent
In the nebular phase, at 175 days after maximum from Tycho Brahe's records,
the SN went back to a white color. Such behaviour, once the correction for
extinction is made, is consistent, once more, with that observed in normal
SNe Ia. This takes into account uncertainties in the color estimates and
includes the new records above.

\smallskip
\noindent
In R04 the visual absolute magnitude of Tycho's SN is
estimated as:

\smallskip
\noindent
$M_{V}$ = -17.72 -5 log($d$/3.5 kpc) - $A_{V}$ mag.
If corrected for $A_{V}$ = 1.86 $\pm$ 0.12 mag, that gives:
$M_{V}$ = -19.58 -5 log($d$/3.5 kpc) $\pm$ 0.42 mag.

\smallskip
\noindent
De Vaucouleurs (1985) gives as the most likely estimate of the distance to
Tycho Brahe's SN: $d$ = 3.2 $\pm$ 0.3 kpc. Adopting 3.2 kpc for the distance,
we have an absolute visual magnitude $M_{V}$ = -19.38 $\pm$ 0.42 mag.
More recently, however, new discussions of the distance sets it to
$d$ = 2.7 $\pm$ 1 kpc. This estimate comes from the growth of extinction
towards higher distance in the field. as measured from stars in
the {\it Gaia DR2} which have measured stellar parameters and colours.
From those
stars one can track the reddening versus distance towards the direction
of SN 1572.  It is consistent with other estimates mentioned later on.
The absolute magnitude of SN 1572 at
maximum should be $M_{V}$ = -19.02 - 5 log($d$/2.7) $\pm$ 0.42 mag, which
compares well with $M_{V}$ = -19.12 $\pm$ 0.26 mag, the mean magnitude
from the Calan Tololo sample (Hamuy et al. 1996a).

\noindent
\section{\bf On the explosion mechanism and progenitor of Tycho's SN} 
\begin{figure}[h!]
\centering
\includegraphics[width=0.8\columnwidth]{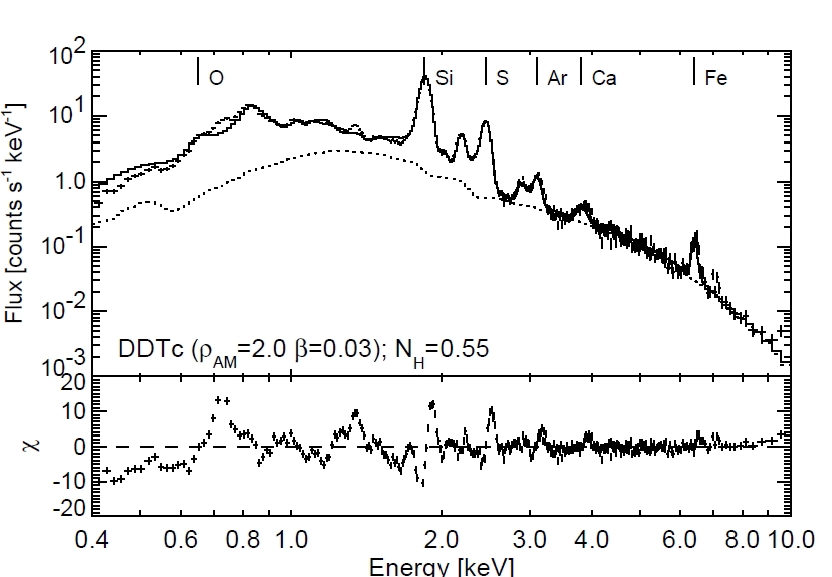}
\caption{Comparison between the emission of the
  ejecta from a delayed detonation model
  (Figure 7 of Badenes et al. 2006)
  and the spatially integrated spectrum of region B.
  Courtesy of C. Badenes. $\copyright$ AAS. Reproduced wih permission.}
\label{fig:5}
\end{figure}

\begin{figure}[h!]
\centering
\includegraphics[width=0.6\columnwidth]{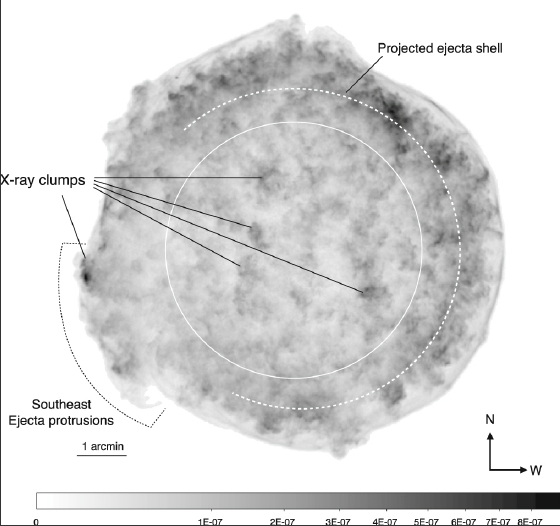}
\caption{
 Figure 2 in Sato et al. (2019). 
  Flux image (1.76–4.2 keV) of Tycho’s SNR observed in 2009 by
  Chandra. The central region (inner white circle)
  of the remnant has been used for
  the genus statistics. The southeast ejecta protusions 
  can be seen as well in the Figure.
  Courtesy of J.P. Hughes. $\copyright$ AAS. Reproduced with permission. }
\label{fig:6}
\end{figure}

\noindent
After Baade (1938, 1945a,b) identified the ``nova'' B Cas, based on the
light-curve data in Tycho Brahe's {\it Progymnasmata}, as a Type I supernova, he
remarked (1945a,b) that no expanding shell had been detected at the position
where the supernova flared up.

\noindent
Exploration of the sky at radio wavelengths, about a decade later (Hanbury
Brown
\& Hazard 1952; Shakeshaft et al. 1955; Baldwin \& Edge 1957) did locate such
nebula at positions compatible with that given by Baade, within the
observational errors. Studies of Tycho's SNR at all wavengths have
followed, now covering the full range from radio waves to $\gamma$-rays. 

\noindent
Especially significant have been the X-ray observations made with the
$ROSAT$ (Hughes 2000), {\it XMM-Newton} (Decourchelle et al. 2001) and
$Chandra$
(Hwang et al. 2002) satellites. Namely, from the {\it XMM-Newton}
data (Decourchelle et al. 2001), it was found that the chemical abundances and
their distribution inside the SNR were those expected from a SN Ia. From the
good correlation between the images in the Si XIII K line and the Fe XVII L
line, they deduced that some fraction of the inner iron layer had been well
mixed with the outer silicon layer.

\begin{figure}[h!]
\centering
\includegraphics[width=0.6\columnwidth]{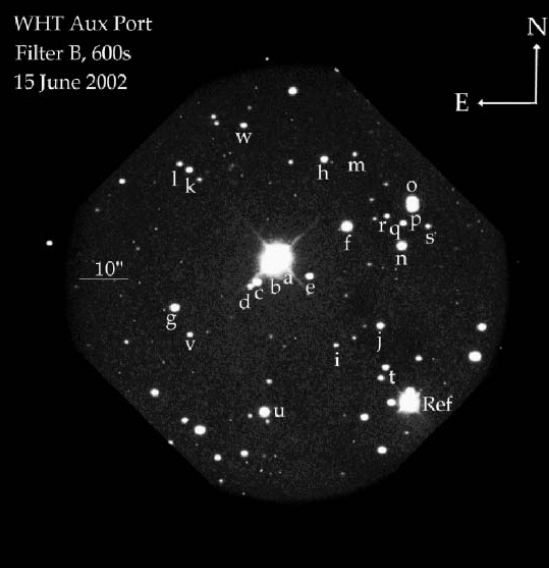}
\caption{Image from the Auxiliary
  Port at the William Herschel Telescope of the center of the field of SN 1572.
  It reveals that it is far from being a crowded field. 
  The initial search area in RL04
  covers a radius of 0.65 arcmin around RA $=$ 00 h 25 min 19.9 s,
  Dec. $=$ 64$^{o}$
  08' 18.2'' (J2000) (the Chandra geometrical centre of X–ray emission). From
  the author's personal archive.}
\label{fig:7}
\end{figure}

\noindent
Based on the increasingly detailed data, at all wavelengths, on the morphology,
dynamics and chemical composition of Tycho's SNR, there has been physical
modeling to infer the explosion mechanism and the nature of the
progenitor system of the SN.

\noindent
Badenes et al. (2006), used X-ray observations from $XMM-Newton$ and $Chandra$
to test different explosion models and found that the fundamental properties
of the X-ray emission in Tycho were well reproduced by a one-dimensional
delayed detonation model of a Chandrasekhar-mass white dwarf (see Figure 5),
interacting with
an ambient medium of density $\rho_{AM} = 2\times10^{-24} g\ cm^{-3}$. There is
stratification of the chemical composition, which points to a supersonic
burning front. 
Kozlova \& Blinnikov (2018), from hydrodynamical calculations of the X-ray
spectrum of Tycho's SNR and comparison of the model with {\it Chandra}
high-resolution images, also conclude that a delayed-detonation model expanding
into a uniform interstellar medium is the preferred one for this SNR.
 The fit of this model gives as best distance 2.8 $\pm$ 0.4 kpc. 
Earlier, Williams et al. (2017), from 3D measurements of the velocities of
various ejec{ta knots using {\it Chandra} X-ray observations over a 12 yr
baseline, had equally found that delayed-detonation models are favored.

\noindent
Badenes et al. (2007), by comparing their models for the evolution of the
SNRs with the observations of several remnants of the Ia type, Tycho's SNR
among them, find incompatibility with the  pre-supernova
models where there is emission of strong, optically thick winds from the
progenitor system of the supernova. Such winds would excavate large,
low-density cavities around the progenitors and that would be incompatible
with the dynamics and the X-ray emission of these SNRs.
 Chiotellis et al. (2013) find that an uniform ambient density
  cannot simultaneously reproduce the dynamical and X-ray emission properties
  of Tycho. A better fit is provided by models in which the remnant was
  evolving within a dense but small wind bubble. The wind bubble might have
  different origins, including a sequence of nova explosions or a
  double-degenerate origin.

\noindent  
Sato et al (2019) analyse the clumpy structure of Tycho's SN remnant.
Their genus statistic analysis supports a scenario in which
the observed structure
of the SN Ia remnant arises from initial clumpiness in the explosion (see
Figure 6).
At present, the cause of the initial ejecta clumping in SNe Ia
is found by these authors to be
still theoretically unclear and they expect it to
gradually become more clear with
a 3D simulation covering from the explosion to the remnant phase.
The most recent 3D study of the velocity of 59 clumpy, metal-rich
ejecta knots, together with the proper motions, estimate a new expansion center
of the SNR, at $\sim$6 arcsecs from the geometrical center
(Millard et al. 2022). These authors also find that
the southeast quadrant expands faster than the rest of the SNR, therefore
confirming some degree of asymmetry in the expansion of the ejecta.

\bigskip 

\noindent
Yamaguchi et al. (2017), adopting again a delayed detonation model for a
Chandrasekhar-mass white dwarf, explain an
iron-rich knot located along the eastern rim of the remnant, surprisingly
with no emission from Cr, Mn or Ni (which implies mass ratios
$M_{Cr}/M_{Fe} < 0.023$, $M_{Mn}/M_{Fe} < 0.012$ and $M_{sNi}/M_{Fe} < 0.029$) as
originating from a region having reached peak temperatures of
(5.3-5.7) $\times 10^{9}$ K only, with a neutron excess $\lapprox 2.0 \times
10^{-3}$, which corresponds either to incomplete Si burning or to an
$\alpha$-rich freeze-out regime, which excludes the dense core of the
white dwarf as its origin and points to a region near
the boundary of the core as the site of production of the knot.

\noindent
As we have seen, the delayed detonation mechanism in a Chandrasekhar-mass
white dwarf appears to be the one that best
explains the overall characteristics of Tycho's SNR in the analysis done
by various authors. In the delayed detonation models, C is ignited
close to the center of the
white dwarf when, due to the accretion of mass from a close binary companion,
the central density reaches $\simeq 2\times10^{9}$ g cm$^{-3}$ and the
temperature rises to $\sim 10^{9}$ K.
Burning then propagates subsonically (a deflagration) outwards, causing the
layers ahead of the burning front to expand. Hydrodynamical instabilities make
the burning front turbulent. When some fraction of the star has already been
burned and the front reaches regions having densities below some critical
value, the front becomes supersonic (a detonation), sweeping the rest of the
star up to the surface.

\smallskip
\noindent
Growth of a C+O white dwarf up to the Chandrasekhar mass by accretion of
material
from a companion star in a binary system results from a comparatively
slow mass transfer (capture of material from the stellar wind of the
companion or from a stream due to Roche-lobe overflow by the mass-donor star).
 Faster mass transfer would happen in the merging with another C+O white dwarf,
although it should not be too fast to avoid C ignition close to the surface of
the accreting white dwarf before the Chandrasekhar mass is reached. If the
explosion would immediately follow the merging, it would hardly have
the characteristics of most SNe Ia. If there were a sufficient  delay between
merging and explosion (MED, see Socker 2019a,b; 2022), however, the exploding object
might have becomed a Chandrasekhar-mass white dwarf and then undergo a
delayed detonation.
Recent work (Neopane et al. 2022) has shown that double-degenerate mergings
can actually produce highly magnetized, uniformly rotating white dwarfs, a
fraction of them with masses close to the Chandrasekhar mass, which should
then explode via the delayed-detonation mechanism.
 Then, although delayed detonation models seem consistent with a single
  degenerate
  path to explosion, they can occur as well, if a sufficient delay between the merging of two white dwarfs and the explosion takes place,
   in the double degenerate scenario.

  \smallskip
  \noindent
  In fact this DD-MED mechanism is the equivalent in the DD case to the spin up/spin down
  from Di Stefano et al (2011) applied with the accretion from a non degenerate
  star (SD path). The SD spin up/spin down models can leave a very
  faint companion.
  Such faint companion would be too faint to be catalogued 
  in the {\it Gaia} data releases or in the {\it Hubble Space Telescope}
  images taken of
  the field thus far.

\noindent
The search for a surviving companion of SN 1572 has been
a survey of the stars located not too far from the present geometrical centroid
of the remnant, at distances within the estimates for that of the SNR, looking
for unusually high tangential and radial velocities, photometric and
spectroscopic peculiarities and possible excess of Fe-peak elements at the
surface (Ruiz-Lapuente 1997; RL04).
\noindent
In the
single degenerate channel,  
the surviving companions of the explosion can, in principle,
be at any stage of thermonuclear
evolution:  main sequence, subgiant, giant or supergiant stars  (see Wang \&
Han 2012; Maoz et al. 2014; Ruiz-Lapuente 2014, 2019,  Soker 2019a
for reviews).
They could also be hot subdwarfs (Meng \& Li 2019; Meng \& Luo 2021) or fainter
objects as mentioned above (Di Stefano et al. 2011). 
Hydrodynamic simulations of the impact of the SNe Ia ejecta on a non-degenerate
companion of any type predict that such stars will survive
the explosion
after being stripped of some of their mass, heated, and their surfaces possibly
contaminated by the slowest moving SN ejecta. The binary system being disrupted,
the companions should be ejected at their orbital velocities, plus some kick
from the impact of the SN material. In the double degenerate scenario
there will be no surviving companion. 

\noindent
The first survey (RL04) (see Figure 7)
included a number of stars smaller than 
the ones made later on. Spectra and
photometry of all the stars were obtained with several telescopes. Images were
taken with the {\it Hubble Space Telescope}. 
Radial velocities were measured from the spectra, and proper motions from
the {\it HST} images. Spectral types and luminosity classes were determined
by modeling the spectra. Comparison with the photometry then gave the distances
to the sampled stars. One star, a subgiant, was a 3$\sigma$
outlier in radial velocity, as compared with the stars at the same distance and
position on the sky. The star, labelled Tycho G (see Fgure 7), was an
outlier in proper motion as well, and
its metallicity showed that it was not a halo star. It was therefore proposed, 
based on its kinematics, as the likely surviving companion of SN 1572.

\smallskip
\noindent
High-resolution spectra of the star above were later obtained with
the {\it HIRES} spectrometer on the {\it Keck I} telescope.
That allowed an
accurate determination of the chemical abundances, in addition to
a refinement of the stellar atmosphere parameters and a more precise
measurement of the radial velocity.
Low-resolution spectra of other stars in the Tycho field were also analyzed to
determine their spectral types, with good agreement with the results of
RL04.

\smallskip
\noindent
Kerzendorf et al. (2009) had remarked that if a companion star were
rotating synchronously with the orbital motion (rotation period equal to the
orbital period), it  should be rotating faster than the proposed
candidate. Gonz\'alez Hern\'andez et al. (2009) had already argued that the
interaction with the SN ejecta can slow down the rotation. Liu et al.
(2013) and Pan et al. (2014) later showed, from hydrodynamical
simulations, that the rotational velocity can indeed be much reduced by the
collision.

\smallskip
\noindent
A number of works have been devoted to the possible companions
of Tycho's SN remnant. 
In Kerzendorf et al. (2013), attention was paid to a fast-rotating star of the
spectral A type, located close to the geometrical center of the SNR,
labelled Tycho B (see Figure 7). 
  Kerzendorf et al. (2018), from UV spectra obtained with the {\it Hubble Space
    Telescope}, concluded that it rather is a foreground star.
  The idea was that if there were enough Fe II 
in the SNR and the star were inside it, the spectrum would show blueshifted absorption lines. 
If it were behind the SNR, in the background,
the lines would be both blueshifted and 
redshifted, as it is the case with the Schweizer-Middleditch star,
in the background of 
SN 1006. No Fe II absorption lines are seen in the spectrum of star B, which by having a 
surface temperature $T_{\rm eff} \sim$ 10,000 K, should have significant UV emission. From 
that, Kerzendorf et al. (2018) conclude that either star B is
in the foreground or there is 
not enough Fe II in the SNR, the material being more highly ionized.
{\it Gaia DR3} parallaxes place Tycho B well in the range of distance
of the SNR.
Ihara et al. (2007) suggested that Tycho E could be the companion of SN 1572
on grounds of absorption seen in the spectra. But this star clearly seems
behind the Tycho Brahe's SN (see the Appendix). As we mention later, both
Tycho B and Tycho E orbits do not look perturbed by an impact.

\noindent  
Bedin et al. (2014) measured the proper motions from {\it HST} images 
of a large sample of stars
around the center of Tycho's SNR. The chemical abundances of the candidate
star Tycho G were calculated in Gonz\'alez Hern\'andez et al (2009), Kerzendorf
et al. (2013) and Bedin et al, (2014).
The Ni/Fe ratio would point, at most, at moderate or low
pollution of the companion
of the SN. However, that  is  what is predicted by the models of Pan et al. (2014): the captured
material would be much diluted in the convective envelope of a subgiant star.

\smallskip
\noindent
 The advent of the {\it Gaia DR2}
opened a new space for the definitive exploration of
the candidate stars to companion of Tycho Brahe's  SN. 
Ruiz-Lapuente et al. (2019) used the parallaxes from
the {\it Gaia DR2}  to reassess the distances to the stars in the
Tycho field. The orbits of the stars were also calculated. No orbital
peculiarity is seen in Tycho B nor in
other candidates stars that had been proposed. The
star in RL04 is the most dynamically peculiar in the
sample, but there is no real proof that it is the surviving companion it has
been looked for.
In order to do a full update of this research, in this review the same
analysis using now the most recent {\it Gaia} data is done,
those from
the {\it Gaia DR3}. The results, which are shown in the Appendix,
do not change the conclusions from
the 2019 paper using {\it Gaia DR2}. The proper motions of the stars
are similar (though not identical) to those from {\it Gaia DR2}. The
parallaxes are better determined. 
Neither Tycho B nor Tycho E have
a peculiar orbit. The only star somehow eccentric
and with a proper motion and V$_{R}$ higher than the rest is Tycho G.
The peculiarity of a star, Tycho G,
comes from three facts:
a larger proper motion mostly in declination than the other stars, a larger
eccentricity and larger radial velocity in the LSR $V_{R}$ (other star, Tycho U,
has a similar proper motion in declination than Tycho G, but no eccentricity and a small V$_{R}$ ). No evidence for
a high chemical pollution from the SN Ia is found in this only relatively
peculiar star. 
It might be that this is
what occurs for an impact on a main sequence or subgiant companion. It should be addressed as well if a star close
to the explosion, but not being the companion, could be perturbed
in the way Tycho G appears to have been.  The single degenerate path to
explosion has not been fully  excluded for Tycho Brahe's supernova, from
all the abovementioned factors.

\bigskip

\noindent
 From the theoretical work
 done on SN impacts on subgiants and on
 main sequence stars (see reviews in the previous paragraphs),
 the velocity of the companion and its pollution might not be very high. 
In the case of a donor star orbiting
very close to the WD, though,  one
would expect an outlier at many more than 3$\sigma$ of the proper motion
distribution.

\smallskip
\noindent
There have been some other observations pointing indirectly to a single
degenerate origin rather than to a double degenerate origin for SN 1572.
Zhou et al. (2016) have found that Tycho's SNR is surrounded by a clumpy
molecular bubble, expanding at $\sim$ 60 km s$^{-1}$. The bubble is massive
and there is morphological correspondence with the SNR. The authors
suggests that
the origin of the expanding bubble is a fast outflow coming fron the vecinity
of the mass-accreting WD that gave rise to SN 1572.

\bigskip
\noindent
    To complete this examination, we would like to see whether there are hypervelocity stars in the field of Tycho's SN remnant.
The generous assumption of a 1 kpc distance to the remnant, 3000 km s$^{-1}$ 
and an age of 450 yr, gives a  search radius of 4.7 arcmin.
No hypervelocity star is found.  But,
      the whole field is totally empty of
hypervelocity stars, even if we amplify the search to a 1 degree around the
geomerical center of the remnant. For such radius  and 
star distances between 1.7 and 3.7 kpc, the sample contains 58,691 stars.
Thus we can exclude as an explosion mechanism for Tycho Brahe's SN the
{\it dynamically driven double–degenerate,
double–detonation scenario
D$^{6}$} mechanism (Shen et al. 2018).
These explosions are
triggered by the detonation of a surface layer made of He,
accreted by the exploding WD from a less massive WD companion.
The outburst
might happen when the mass–donor has not yet been tidally disrupted. Due to its
very high orbital velocity,
the WD companion should be ejected as a hypervelocity
star (v $>$ 1000 km s$^{-1}$ ). As we have seen, this is not the case for
SN 1572.

\begin{figure}[h!]
\centering
\includegraphics[width=0.6\columnwidth]{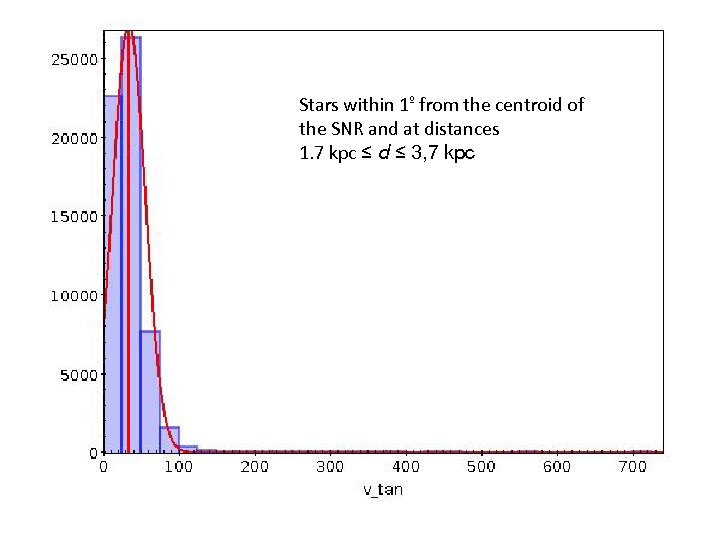}
\caption{ Histogram of the distribution of tangential velocities of
  the stars within 1 degree of the geometrical center of
  Tycho’s SNR in the range of distance
  compatible with SN 1572 (1.7 $<$ d $<$ 3.7 kpc). The data are obtained
  from {\it Gaia DR3}. It can be seen that there are no hypervelocity stars
in a very large area
around the Tycho's SN remnant.} 
\label{fig:8}
\end{figure}

\noindent
In conclusion of this section,  
the explosion mechanism favored by the analysis of various authors suggests
that a delayed detonation explains better the data than other possible
alternatives. The single degenerate and the double degenerate scenario with
a delay between merging and explosion fit into this
picture.   The new scrutiny using {\it Gaia DR3} of the stars in the field
of  SN 1572 gives a similar conclusion to that reached with the {\it
  Gaia DR2}. A few interesting notes and Figures can be found in he
Appendix. 

\smallskip

\noindent
In the next decade, we expect to probe the nature of SNe Ia and quantify those
coming from mergings of WDs with the new generation of gravitational
wave missions in the  deci-hertz range (see Yoshida (2020), for instance). We would
be able to quantify the rates of SNe Ia coming through this path versus
other origins. In the meanwhile,
we have a battery of tests that are not so robust
as a direct detection. 

\noindent
    In the conclusions, I summarize the results gathered on the origin
      of the eariest SN Ia observed by astronomers.

\section{Conclusions} 

\noindent
We found in R04 that SN 1572 was a supernova very close to the template
with a {\it stretch factor}  s $\sim$ 0.9. 
The light curve grows in precision towards the late 
times, being highly uncertain around maximum brightness.
An overall agreement between early, late decline and color
with the expected evolution of normal SNeIa 
supports our conclusion.

\noindent
Type Ia supernovae with {\it stretch factors} between 0.9 and 1.1 
make the vast majority of the observed population. They are not only
those most frequently found in nearby searches, but also the bulk
of discoveries in cosmological searches at high--z, as can be seen in
the sample of SNe at z $>$ 0.3 found by the 
Supernova Cosmology Project (Perlmutter et al. 1999). Among SNe Ia
of s $\sim$ 0.9 in nearby galaxies 
for which very late--time data are available, we have  
found a close resemblance to SN 1996X in rate 
of decline. SN 1572 likely has a slightly slower rate. However, whereas 
SN 1996X was not heavily reddened, the reddening in SN 1572 is
E(B--V)=0.60 $\pm$ 0.05.  The echo of SN 1572 discovered in 2008,
reaffirms that SN 1572 is a normal SN Ia.

\noindent
X-ray observations of Tycho's SNR have provided an increasingly detailed
  picture of the remnant and its surroundings. Although one-dimensional models
  do fit the overall charateristics of the SNR, three-dimensional simulations
  of its evolution, from the explosion to its current state, are being
  developed to account for the detailed structure.

\noindent
Concerning the mechanism of explosion, a delayed detonation model
seems to give a better account of the nucleosynthesis, as derived
by various authors.

\noindent
 The possibility for a surviving companion has been
throughly reexamined here with {\it Gaia DR3} data. The conclusion is similar
to the one reached with the {\it Gaia DR2}. 

\bigskip

\noindent
Something seems clear 
from the exploration of the field of Tycho Brahe's supernova:
the explosion is not 
triggered by the detonation of a surface layer made of He,
accreted by the exploding WD from a less massive WD companion. Thus,
it does not come
from the so--called  {\it dynamically driven double–degenerate,
  double–detonation scenario}. There are no hypervelocity stars
in the field of SN 1572. 

\noindent
Finally,
the merging of two WDs with a lapse between merging and explosion presents
a possible path to this normal SN Ia: hydrodynamical calculations,
which show that a delayed detonation in a near Chandrasekhar mass 
WD can occur in a double degenerate scenario where the explosion
occurs with a delay after merging, change the view on the DD scenario. 
The delayed detonation of a Chandrasekhar WD had been for long
investigated through
the growth of the WD by accretion from a non--degenerate companion.
But both paths seem possible for normal SNe Ia. Further exploration
is needed to clarify the case of SN 1572. 

\noindent
Several aspects on Tycho Brahe's supernova are still unsolved  450 yrs
after its visual detection. Being a normal SN Ia what can be learnt from this
explosion, can help to understand the wide majority of supernovae of this
type.

\noindent{\bf Acknowledgements.} 
I thank Ralph Neuhaeuser for sharing his findings presented at the EAS in 2022
with me. I would like to thank Jack Hughes, Carles Badenes and Tomonori Usuda
for their kind permission to show Figures from their Tycho's SN papers.
 I would like to thank Bob Fisher for comments on Chandrasekhar explosions
through the double degenerate channel and  two anonymous referees for pointing
out relevant aspects to Tycho Brahe's SN understanding. 
This work is supported by  the project PID2021-123528NB-I00, from the
Ministerio de Ciencia e Innovaci\'on of Spain.

\noindent
 This work has made extensive use of the {\it Gaia} DR3.
{\it Gaia} data are being processed by
the {\it Gaia} Data Processing and Analysis Consortium (DPAC).
Funding for the DPAC is provided by national institutions, in
particular the institutions participating in the {\it Gaia} MultiLateral Agreement (MLA). The Gaia mission website is https:
//www.cosmos.esa.int/gaia. The {\it Gaia} archive website is
https://archives.esac.esa.int/gaia.

\vfill\eject

%\begin{thebibliography}{}

\noindent
{\bf References}

\noindent
Baade, W. (1938). The absolute photographic magnitude of supernovae. {\it
Contr. Mnt. Wilson Obs./Carnegie Inst. Was.}, 600, 1-20

\noindent
Baade, W. (1945a). B Cassiopeiae as a supernova of Type I. {\it Contr.
Mnt. Wilson Obs./Carnegie Inst. Was.}, 711, 1-9  

\noindent 
Baade, W. B. (1945b). B Cassiopeiae as a Supernova of Type I. {\it Astrophys.
J.}, 102, 309. doi:10.1086/144761 

\noindent  
Badenes, C., Borkowski, K.J., Hughes, J.P., Hwang, U., and  Bravo, E. (2006).
Constraints on the Physics of Type Ia Supernovae from the X-Ray Spectrum of
the Tycho Supernova Remnant. {\it  Astrophys. J.}, 645, 1373-1391.
doi:10.1086/504399 

\noindent
Badenes, C., Hughes, J.P., Bravo, E., and Langer, N. (2007), Are the Models
for Type Ia Supernova Progenitors Consistent with the Properties of Supernova
Remnants? {\it  Astrophys. J.}, 662, 472-486, doi:10.1086/518022 

\noindent
Baldwin, J.E., and Edge, D.O. (1957). Radio emission from the remnants of
the supernovae of 1572 and 1604. {\it Observatory}, 77, 139-143

\noindent
Bauer, E.B., White, C.J., and Bildsten, L. (2019). Remnants of Subdwarf Helium
Donor Stars Ejected from Close Binaries with Thermonuclear Supernovae. {\it
Astrophys. J.}, 887, 68. doi:10.3847/1538-4357/ab4ea4

\noindent
Bedin, L.R., Ruiz-Lapuente, P., Gonz\'alez Hern\'andez, J.I., et al. (2014). 
Improved Hubble Space Telescope proper motions for Tycho-G and other stars in
the remnant of Tycho's Supernova 1572. {\it Month. Not. Roy. Astron. Soc.}, 439,
354-371. doi:10.1093/mnras/stt2460
  
\noindent
Brahe,T. (1603a). Astronomiae Instauratae Progymnasmata, in
{\it Opera omnia} 2, 307, ed. I.L.E. Dreyer (Amsterdam: Swets \& Zeitlinger
1972)

\noindent
Brahe,T. (1603b>). Astronomiae Instauratae Progymnasmata, in
{\it Opera omnia} 3, 81, ed. I.L.E. Dreyer (Amsterdam: Swets \& Zeitlinger 1972)

\noindent
Cappellaro, E., Patat, F., Mazzali, P. A., Benetti, S., Danziger,
I.J., Pastorello, A., et al. (2001). Detection of a Light Echo from SN 1998BU.
{\it Astrophys. J.}, 549, L215-L218. doi: 10.1086/319178

\noindent
 Chiotellis, A., Kosenko, D., Schure, K.M., Vink, J., and Kaastra, J.S.
(2013). Modelling the interaction of thermonuclear supernova remnants
with circumstellar structures: the case of Tycho's supernova remnant.
{\it Month. Not. Roy. Astron. Soc.}, 435, 1659. doi:10.1093/mnras/stt1406

\noindent
  Decourchelle, A., Suavageot, J.L., Audard, M., Aschenbach, B,
  Sembay, S., Rothenflug, R., et al. (2001). XMM-Newton observation of the
  Tycho supernova remnant. {\it Astron. Astrophys.}, 365,
  L218-L224. dpi:10.1051/0004-6361:20000115

\noindent
  de Vaucouleurs, G. (1985). Tycho's supernova and the Hubble constant.
  {\it Astrophys. J.}, 289, 5-9. doi:10.1086/162858

  \noindent
  Di Stefano, R., Voss, R. Claeys, J.S.W.(2011).
  Spin-up/Spin-down models for Type Ia supernovae. {\it Astrophys.. J. Lett.},
  738, L1, 1-9 (2011). doi:, 10.1088/2041-8205/738/1/L1

\noindent
  Drimmel, R., and Spergel, D.N. (2001). Three-dimensional Structure of the
  Milky Way Disk: The Distribution of Stars and Dust beyond 0.35 R$_{solar}$.
  {\it  Astrophys. J}, 556, 181-202. doi:10.1086/321556  
 
\noindent
  Filippenko, A.V., Richmond, M.W., Branch, D., Gaskell, M., Herbst, W-, Ford,
  C.H., et al. (1992). The Subluminous, Spectroscopically Peculiar
  Type 1a Supernova 1991bg in the Elliptical Galaxy NGC 4374. {\it Astron. J.},
  104, 1543-1556. doi:10.1086/116339
  
\noindent
  Goldhaber, G., Groom, D.E., Kim, A., Aldering, G., Astier, P., Conley, A.,
  et al. (2001). Timescale Stretch Parameterization of Type Ia
  Supernova B-Band Light Curves. {\it Astrophys. J.}, 558, 359-368.
  doi:10.1086/322460 

\noindent
  Gonz\'alez Hern\'andez, J. I., Ruiz-Lapuente, P., Filippenko, A. V.,
  Foley, R.J., Gal-Yam, A., and Simon, J.D. (2009). The Chemical Abundances of
  Tycho G in Supernova Remnant 1572. {\it Astrophys. J.}, 691, 1-15.
  doi:10.1088/0004-637X/691/1/1

\noindent
  Guy, J., Astier, P., Baumont, S., Hardin, D., Pain, R., Regnault, N., et al.
  (2007). SALT2: using distant
  supernovae to improve the use of type Ia supernovae as distance indicators.
  {\it Astron. Astrophys.}, 466, 11-21. doi:10.1051/0004-6361:20066930

\noindent
Hamuy, M., Phillips, M.M., Schommer, R.A., Suntzeff, N.B., Maza, J., and 
Avil\'es, R. (1996a). The Absolute Luminosities of the Calan/Tololo Type IA
Supernovae. {\it Astron. J.}, 112, 2391-2397. doi:10.1086/118190 

\noindent
  Hamuy, M., Phillips, M.M., Suntzeff, N.B., Schommer, R.A., Maza, J., Smith,
  R.C., et al. (1996b). The Morphology of Type IA Supernovae Light Curves
  {\it Astron. J.}, 112, 2438-2447. doi:10.1086/118193

\noindent
  Hanbury Brown, R., and Hazard, C. (1952). Radio-Frequency Radiation from Tycho
  Brahe's Supernova (A.D. 1572). {\it Nature}, 170, 364-366.
  doi:10.1038/170364a0 
 
\noindent
  Hernandez, M., Meikle, W.P.S., Aparicio, A., Benn, C.R., Burleigh, M.,
  Chrysostomou, A.C., et al. (2000). An early-time infrared and optical study
  of the Type Ia Supernova 1998bu in M96. {\it Month. Not. Roy. Astron. Soc.},
  319, 223-234. doi:10.1046/j.1365-8711.2000.03841.x

\noindent
  Hughes, J.P. (2000). The Expansion of the X-ray
  Remnant of Tycho's Supernova (SN 1572), {\it Astroph. J. Lett.},
 545, L53-L56. doi:10.1086/317337

\noindent
Hughes, J.P., Hayashi, I., Helfand, D., Hwang, U., Itoh, M., 
Kirshner, R., et al. (1995). ASCA Observations of the Large Magellanic Cloud
Supernova Remnant Sample: Typing Supernovae from Their Remnants.
{\it Astrophys. J. Lett}, 444, L81-L84. doi:10.1086/187865 

\noindent
  Hwang, U, Petre, R., Szymkowiak, A.E., and Holt, S.S. (2002). Chandra
  Observations of Tycho's Supernova Remnant. {\it Jour. Astrophis. Astron},
  23, 81-87. doi:10.1007/BF02702469

\noindent
  Ihara, Y., Ozaki, J., Doi, M., Shigeyama, T.,
 Kashikawa, N., Komiyama, Y. and Hattori, T. (2007).
 Searching for a Companion Star of Tycho's Type Ia Supernova with Optical
 Spectroscopic Observations. {\it Publ. Astron. Soc. Japan}, 59, 811-825.
 doi:10.1093/pasj/59.4.811

\noindent
  Jha, S., Riess A.G., and Kirshner, R.P. (2007). Improved Distances to Type Ia
  Supernovae with Multicolor Light-Curve Shapes: MLCS2k2. {\it Astrophys. J.},
  659, 122-148. doi:10.1086/512054 

\noindent
  Katsuda, S., Petre, R., Hughes, J.P., Hwang, U., Yamaguchi, H., Hayato, A.,
  et al. (2010). X-ray Measured Dynamics of Tycho's Supernova Remnant.
  {\it Astrophys. J.}, 709, 1387-1395. doi:10.1088/0004-637X/709/2/1387 
 
\noindent
  Kerzendorf, W.E., Schmidt, B.P., Asplund, M., Nomoto, K., Podsiadlowski,
  Ph., Frebel, A., et al. (2009). Subaru High-Resolution Spectroscopy of Star
  G in the Tycho Supernova Remnant. {\it Astrophys. J.}, 701, 1665.1672.
  doi:10.1088/0004-637X/701/2/1665

\noindent
  Kerzendorf, W. E., Yong, D., Schmidt, B.P., Simon, J.D., Jeffery, C.S.,
  Anderson, J., et al. (2013). A High-resolution Spectroscopic Search for the
  Remaining Donor for Tycho's Supernova. {\it Astrophys. J.}, 774, 99 (K13).
  doi:10.1088/0004-637X/774/2/99

\noindent
  Kerzendorf, W.E., Long, K.S., Winkler, P.F., and Do, T. (2018). Tycho-B: an
  unlikely companion for SN 1572.
  {\it Month. Not. Roy. Astron. Soc.}, 479, 5696-5703.
  doi:10.1093/mnras/sty1863

\noindent
  Kozlova, A.V., and Blinnikov, S.I. (2018). Distance Estimate of Tycho’s SNR.
  {\it Jour. Phys.Conf. Ser.}, 1038, 012006.
  doi:10.1088/1742-6596/1038/1/012006 

\noindent
  Knop, R.A, Aldering, G., Amanullah, R., Astier, P., Blanc, G., Burns, M.S.,
  et al. (2003) (SCP2003). New Constraints on $\Omega_{M}$, $\Omega_{\Lambda}$,
  and $w$ from an Independent Set of 11 High-Redshift Supernovae Observed with
  the Hubble  Space Telescope. {\it Astrophys. J.}, 598, 102-137.
  doi:10.1086/378560

\noindent
 Kozlova, A.V., and Blinnikov, S.I. (2018). Distance Estimate of Tycho's
  SNR. {\it IOP Conf. Ser.: Jour. of Phys.: Cof. Ser. 1038, 012006.
    doi:10.1088/1742-6596/1038/1/012006}

\noindent
  Krause, O., Tanaka, M., Usuda, T., Hattori, T., Goto, M., Birkmann, S., et
  al. (2008). Tycho Brahe's 1572 supernova as a standard type Ia as revealed by
  its light-echo spectrum{\it Nature}, 456, 617-619.
  doi:10.1038/nature07608

\noindent
  Leibundgut, B., Kirshner, R.P., Phillips, M.M., Wells, L.A., Suntzeff, N.B.,
  Hamuy, M., et al. (1993). SN 1991bg: A Type IA Supernova With a Difference.
  {\it Astron. J.}, 105, 301-313. doi:10.1086/116427

\noindent
  Li, W., Filippenko, A.V., Treffers, R.R., Riess, A.G., Hu, J., and Qiu, Y.
  (2001). A High Intrinsic Peculiarity Rate among Type IA Supernovae.
  {\it Astrophys. J.}, 546, 734-743. doi:10.1086/318299 

\noindent
  Liu, Z.-W., Pakmor, R., R\"opke, F.K., Edelmann, P., Hillebrandt, W.,
  Kerzendorf, W.E., et al. (2013). Rotation of surviving companion stars after
  type Ia supernova explosions in the WD+MS scenario. {\it Astron. Astrophys.},
  554, A109. doi:10.1051/0004-6361/201220903

\noindent
Lira, P. (1995). Master's Thesis, Univ. of Chile

\noindent
Lynn, W.T. (1883). The star of 1572. {\it Observatory}, 6, 151-152 

\noindent
  Maoz, D., Mannucci, F., \& Nelemans, G. (2014). Observational Clues to the
  Progenitors of Type Ia Supernovae. {\it Ann. Rev. Astron. Astrophys.}, 52,
  107-170. doi:10.1146/annurev-astro-082812-141031

\noindent
  Meng, X., and Li, J. (2019). Subdwarf B stars as possible surviving
  companions in Type Ia supernova remnants {\it Month. Not. Roy. Astron. Soc.},
  482, 5651-5655. doi:10.1093/mnras/sty3092

\noindent
  Meng, X., and Luo, Y.-P. (2021). Hot subdwarfs from the surviving companions
  of the white dwarf + main-sequence channel of Type Ia supernovae. 
  {\it Month. Not. Roy. Astron. Soc.}, 507, 4603-4617.
  doi:10.1093/mnras/stab2369 

\noindent
  Millard, M.J., Park, S., Sato, T., Hughes, J.P., Slane, P.,  Patnaude, D.,
  et al. (2022). The 3D X-ray Ejecta Structure of Tycho's Supernova Remnant.
  (2022). {\it Astrophys. J.}, 947, 121. doi: 10.3847/1538-4357/ac8f30 

\noindent
  Milne, P.A., The, L.-S., and Leising, M.D. (1999). Positron Escape from Type
  IA Supernovae {\it Astrophys. J. Suppl.}, 124, 503-526.
  doi:10.1086/313262 

\noindent
  Morgan, F.P. (1945). Tycho's Nova, 1572. {\it Jour. Roy, Astron. Soc. Canada},
  39, 370-372. 

\noindent
Mu\~noz, J. (1573). Libro del nuevo cometa (Valencia: Pedro de 
Huete), re--ed. V. Navarro--Brotons (Valencia: Hispaniae Scientia  1981)

\noindent
 Neopane, S., Bhargava, K., Fisher, R., Ferrari, M., Yoshida, S., Toonen,
S., et al. (2022). Near-Chandrasekhar-mass Type Ia Supernovae from the
Double-Degenerate Channel. {\it Astrophys. J.}, 925, 92.
doi:10.3847/1538-4357/ac3b52

\noindent
Neuhaeuser, R. (2022). Presented at the 
EAS 2002, European Astronomical Society meeting, 27 June--1 July, 2022.
  Celebrating 450 yrs of Tycho's Nova Stella: the physics of supernova
  remnants. History of the SN 1572 discovery (and its color evolution).   
(private communication). 

\noindent
  Nobili, S., Goobar, A., Knop, R., and Nugent, P. (2003). The intrinsic colour
  dispersion in Type Ia supernovae. {\it Astron. Astrophys.}, 404, 901-912.
  doi:10.1051/0004-6361:20030536 
 
\noindent
  Pan, K.C., Ricker, P.M., and Taam, R.E. (2014). Search for Surviving
  Companions in Type Ia Supernova Remnants. {\it Astrophys. J.}, 792, 71.
  doi:10.1088/0004-637X/792/1/71

\noindent
Perlmutter, S. A., Deustua, S.,  Gabi, S., Goldhaber, G., Groom, D., Hook, I.,
et al. (1997). Scheduled discovery of 7+ high-redshift SNe: first cosmology
results and bounds on $q_{0}$. In {\it Thermonuclear Supernovae}, ed.
P. Ruiz-Lapuente, R.Canal, and J.Isern (Kluwer Acad. Publ., Dordrecht, p.749

\noindent
  Perlmutter, S., Aldering, G., Goldhaber,G., Knop, R.A. Nugent, P., Castro,
  P., et al. (SCP99) (1999). Measurements of $\Omega$ and $\Lambda$
  from 42 High-Redshift Supernovae. {\it Astrophys. J.}, 517, 565-586.
  doi:10.1086/307221

\noindent
  Phillips, M.M. (1993). The Absolute Magnitudes of Type IA Supernovae.
  {\it Astrophys. J. Lett.}, 413, L105-L108.
  doi:10.1086/186970 

\noindent
  Phillips, M.M., Phillips, A.C., Heathcote, S.R., Blanco, V.M., Geisler, D.,
  Hamilton, D., et al. (1987). The type IA supernova 1986G in NGC 5128 :
  optical photometry and spectra. {\it Publ. Astron. Soc. Pacific}, 99,
  592-605. doi:10.1086/132020 

\noindent
Phillips, M.M., Lira, P., Suntzeff, N.B., Schommer, R.A., Hamuy, M., 
and Maza, J. (1999). The Reddening-Free Decline Rate Versus Luminosity
Relationship for Type IA Supernovae. {\it Astron. J.}, 118, 1766-1776 (P99).
doi:10.1086/301032

\noindent
  Pskovskii, Y.P. (1977). Photometric classification and basic parameters of
  type I supernovae  {\it Soviet Astron. Lett.}, 3, 215-216

\noindent
  Pskovskii, Y.P. (1984). Photometric classification and basic parameters of
  type I supernovae  {\it Soviet Astron.}, 28, 658-564

\noindent
  Rieke, G.H., and Lebofsky, M.J. (1985). The interstellar extinction law from 1
  to 13 microns. {\it Astrophys. J.}, 288, 618-621.
  doi: DOI:10.1086/162827
 
\noindent
  Riess, A.G., Press, W.H, and Kirshner, R.P. 1995. Using Type IA Supernova
  Light Curve Shapes to Measure the Hubble Constant. {\it Astrophys. J. Lett.},
  438, L17-L20. doi:10.1086/187704 

\noindent
  Riess, A.G, Press, W.H, and Kirshner, R.P. (1996). A Precise Distance
  Indicator: Type IA Supernova Multicolor Light-Curve Shapes. 
  {\it Astrophys. J.}, 473, 88-109. doi:10.1086/178129   

\noindent
  Riess, A.G., Filippenko, A.V., Challis, P., Clocchiatti, A., Diercks, A.,
  Garnavich, P.M., et al. (1998). Observational Evidence from Supernovae for an
  Accelerating Universe and a Cosmological Constant.
  {\it Astron. J.}, 116, 1009-1038. doi: DOI:10.1086/300499

\noindent
  Ruiz-Lapuente, P. (1997). The quest for a supernova companion. {\it Science}.
  276, 1813-1814. doi:10.1126/science.276.5320.1813

\noindent
Ruiz-Lapuente, P., Jeffery, D.J., Challis, P.M., Filippenko,
A.V., Kirshner, R.P., Ho, L.C., et al. (1993). A possible low-mass type Ia
supernova. {\it Nature}, 365, 728-730. doi:10.1038/365728a0   

\noindent
  Ruiz--Lapuente, P., \& Spruit, H.C. (1998). Bolometric Light Curves of
  Supernovae and Postexplosion Magnetic Fields. {\it Astrophys. J.}, 500,
  360-373. doi:10.1086/305697 

\noindent
Ruiz-Lapuente, P. (2004). Tycho Brahe's Supernova: Light from Centuries Past.
  {\it Astrophys. J.},  612, 357-363 (R04). doi:10.1086/422419 

\noindent
  Ruiz-Lapuente, P., Comer\'on, F., M\'endez, J., Canal, R., Smartt, S.J.,
  Filippenko, A.V., et al. (2004). The binary
  progenitor of Tycho Brahe's 1572 supernova. {\it Nature}, 431, 1069-1072.
  doi:10.1038/nature03006

\noindent
Ruiz-Lapuente, P. (2005). in Astronomy as a Model for the Sciences in
Early Modern Times, ed. B. Fritscher, and A. Kuehe (Augsburg: Erwin Rauner). 

\noindent
  Ruiz-Lapuente, P., Damiani, F., Bedin, L., González Hernández, J.I.,
  Galbany, L., Pritchard, J., et al. (2018). No Surviving Companion in
  Kepler's Supernova. {\it Astrophys. J.}, 862, 124.
  doi:10.3847/1538-4357/aac9c4

\noindent
Ruiz-Lapuente, P. (2014). New approaches to SNe Ia progenitors. 
  {\it New. Astron. Rev.}, 62, 15-21. doi:10.1016/j.newar.2014.08.002  

\noindent
  Ruiz-Lapuente, P. (2019). Surviving companions of Type Ia supernovae: theory
  and observations. {\it New Astron. Rev.}, 85, 101523.
  doi:10.1016/j.newar.2019.101523

\noindent
  Ruiz-Lapuente, P., Gonz\'alez Hern\'andez, J.I., Mor, R., Romero-G\'omez,
  M., Miret-Roig, N., Figueras, F., et al. (2019).
  Tycho's Supernova: The View from Gaia. {\it Astrophys. J.}, 870, 135.
  doi:DOI:10.3847/1538-4357/aaf1c1

\noindent
  Salvo, M., Cappellaro, E., Mazzali, P A., Benetti, S., Danziger, I.J.,
  Patat, F., et al. (2001). The template type Ia supernova 1996X.
  {\it Month. Not. Roy. Astron. Soc.}, 321, 254-268.
  doi:10.1046/j.1365-8711.2001.03995.
 
\noindent
  Sato, T., Hughes, J.P., Williams, B.J., and Mori, M. (2019).
  Genus Statistic Applied to the X-Ray Remnant of SN 1572: Clues to the
  Clumpy Ejecta Structure of Type Ia Supernovae
  {\it Astrophys. J.}, 879, 64. doi:10.3847/1538-4357/ab24db 

\noindent
  Shakeshaft, J.R., Ryle, M., Baldwin, J.E., Elsmore, B., and Thomson, J.H.
  (1955). A survey of radio sources between declinations —38$^{o}$ and
  +83$^{o}$. {\it Mem. Roy. Astron.Soc.}, 67, 106-154 

\noindent
  Schmidt, B. P., Kirshner, R.P., Leibundgut, B., Wells, L.A., Porter, A.C.,
  and Ruiz--Lapuente, P., et al. (1994). SN 1991T: Reflections of Past Glory.
{\it Astrophys. J. Lett.}, 434, L19-21. doi:10.1086/187562  

\noindent
  Shen, K.J., Boubert, D., Gänsicke, B.T., et al. 2018,
Three Hypervelocity White Dwarfs in Gaia DR2: Evidence for Dynamically Driven Double-degenerate Double-detonation Type Ia Supernovae.
  {\it ApJ}, 865, 15. doi:10.3847/1538-4357/aad55b

\noindent 
Sneden, C., Gehrz, R.D., Hackwell, J.A., York, D.G., and Snow, T.P. 
(1978). Infrared color and the diffuse interstellar bands.
{\it Astrophys. J.}, 223, 168-179. doi:10.1086/156247

\noindent
Soker, N. (2019a). Supernovae Ia in 2019 (review): A rising demand for
spherical explosions. {\it New Astro. Rev.}, 87, 101535.
doi:10.1016/j.newar.2020.101535

\noindent
     Soker, N (2019b).
      Common envelope to explosion delay time of Type Ia supernovae. {\it Month. Not. Roy. Astron. Soc.}, 490, 2430-2435. doi:10.1093/mnras/stz2817

      \noindent
      Soker, N. (2022). Common Envelope to Explosion Delay time Distribution
(CEEDTD) of Type Ia Supernovae. {\it Res. Astron. Astrophys.}, 22, 035025.
  doi: 10.1088/1674-4527/ac4d25

\noindent
Tian, W.W., and Leahy, D.A. (2011). Tycho SN 1572: A Naked Ia Supernova
Remnant Wihout An Associated Ambient Molecular Cloud. {\it Astrophys. J.
Lett.}, 729. L15. doi:10.1088/2041-8205/729/2/L15

\noindent
Turatto, M., Benetti, S., Cappellaro, E., Danziger, I.J., 
Della Valle, M., and Gouiffes, C. (1996). The properties of the peculiar type
Ia supernova 1991bg. I. Analysis and discussion of two years of observations.
{\it Month. Not. Roy. Astron. Soc.}, 283, 1-17. doi:10.1093/mnras/283.1.1 

\noindent
  Usuda, T., Krause, O.,  Tanaka, M., Hattori, T., Goto, M., Birkmann, S. M.,
  et al. (2013). Light-Echo Spectrum Reveals the Type of Tycho Brahe's 1572.
  in {\it Supernova. Binary Paths to Type Ia Supernovae Explosions}, Proc.IAU
  Symp., 281. doi:10.1017/S1743921312015323 

\noindent
  van den Bergh, S. (1970). Extra-galactic Distance Scale. {\it Nature}, 225,
  503-505. doi:10.1038/225503a0  

\noindent
  van den Bergh, S. (1993). Was Tycho's Supernova a Subluminous Supernova of
  Type Ia? {\it Astrophys. J.}, 413, 67-69. doi:10.1086/172977 

\noindent
Wang, B., and Han, Z. (2012). Progenitors of type Ia supernovae. 
  {\it New Astron. Rev.}, 56, 122-141. doi:10.1016/j.newar.2012.04.001 

\noindent
  Williams, B.J., Coyle, N.M., Yamaguchi, H., Depasquale, J., Seitenzahl, I.R,
  Hewitt, J.W., et al. (2017).
  The Three-dimensional Expansion of the Ejecta from Tycho's Supernova Remnant. 
 {\it Astrophys. J.}, 842, 28. doi:10.3847/1538-4357/aa7384

\noindent
  Yamaguchi, H., Hughes, J.P., Badenes, C.,  Bravo, E., Seitenzahl, I.R.,
  Martínez-Rodríguez, H., et al. (2017). The Origin of the Iron-rich Knot in
  Tycho’s Supernova Remnant.{\it Astrophys. J.}, 834, 124.
  doi:10.3847/1538-4357/834/2/124

  \noindent
  Yoshida, S. (2020). Decihertz Gravitational Waves from Double White Dwarf Merger Remnants. {\it Astrophys. J.}, 906, 20, 1-10.
doi:10.3847/1538-4357/abc7bd

\noindent
Zhou, P., Chen, Y., Zhang, Z.-Y., Li, X.-D., Safi-Harb, S, Zhou, X., et al.
  (2016).
  Expanding MoleculBar Bubble Surrounding Tycho's Supernova Remnant (SN 1572)
  Observed with the IRAM 30 m Telescope:  Evidence for a Single-degenerate
  Progenitor. {\it Astrophys. J.}, 825, 34. doi:10.3847/0004-637X/826/1/34

  \vfill\eject

  \bigskip

  \noindent{\bf Appendix}

  \bigskip

  \noindent

  The most recent survey for a companion in Tycho Brahe's SN has
  been done so far
  using the {\it Gaia DR2} (Ruiz--Lapuente et al. 2019). With the recent {\it Gaia DR3}, we have
  examined whether those results are still
  valid. This is in fact the case: there are minor differences in
  the proper motions and parallaxes to the stars, but they do not affect
  the qualitative results. In Table 1, we display the differences for
  the stars that were used for the orbital calculations. In Figure 9,
  we present the new orbits of the stars.  In Figure 10 we show where Tycho G
  stands in proper motion in the field of stars
  1 degree around the geometrical
  center of the SNR. It is very similar to what it was found with the
  {\it Gaia DR2}. In Figure 11 we show how the sample stars used in
  previous Figures are located now in the {\it Gaia DR3}. Some stars
  have improved their parallax determination. But the overall picture
  is the same.

\begin{table*}
\centering
  \caption{Proper motions of stars {\it G, B, F} and {\it U}, in
    {\it Gaia} DR2 and DR3}
  \begin{tabular}{lcccc}
    \hline
    \hline
    Star & $\mu_{\alpha}^{*}$ (DR2)& $\mu_{\alpha}^{*}$ (DR3) &$\mu_{\delta}$ (DR2) &
    $\mu_{\delta}$ (DR3) \\
    & (mas yr$^{-1}$) & (mas yr$^{-1}$) & (mas yr$^{-1}$) & (mas yr$^{-1}$) \\
    \hline
G & -4.417$\pm$0.191 & -4.253$\pm$0.093 & -4.164$\pm$0.143 & -4.202$\pm$0.097\\
B & -4.505$\pm$0.063 & -4.201$\pm$0.030 & -0.507$\pm$0.049 & -0.518$\pm$0.031\\
F & -5.739$\pm$0.130 & -5.860$\pm$0.054 & -0.292$\pm$0.097 & -0.273$\pm$0.058\\
U & -1.877$\pm$0.113 & -1.658$\pm$0.054 & -5.096$\pm$0.083 & -4.904$\pm$0.057\\
\hline
\end{tabular}
\end{table*}

\bigskip

\begin{figure}[h!]
\centering
\includegraphics[width=0.6\columnwidth]{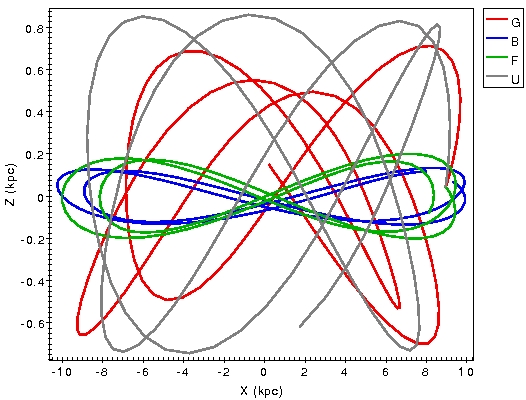}
\includegraphics[width=0.6\columnwidth]{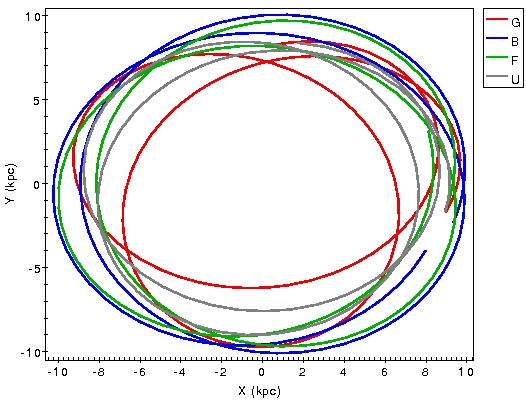}

\caption{ These Figures show the orbits of the stars with the new {\it Gaia DR3} proper motion data. There are no substantial
  changes compared to our 2019 exploration where we used the {\it Gaia DR2}.
The orbits of stars B (green), G (red), F ((blue), and U (gray), projected on the Galactic meridian plane
(up) and on the Galactic plane (bottom), computed forward on time for the next 500 Myr.  In the upper panel,
we see that star U reaches the largest distance from the Galactic plane,
followed by star G, while
stars B and F scarcely depart from the plane. The behaviour of the latter stars is typical of the rest of the sample
considered here. In the bottom panel, we see that the orbit of star G, on the Galactic plane, is quite eccentrical, which
corresponds to the high value of the proper motion and V$_{R}$
compared to the sample, while the other stars (including star U)
have orbits close to circular. Also here, the behaviour of stars B and F is representative of the whole sample. Star E has a similar orbit than stars B and F.
}
\label{fig:9}
\end{figure}

\bigskip

\begin{figure}[h!]
\centering
\includegraphics[width=0.6\columnwidth]{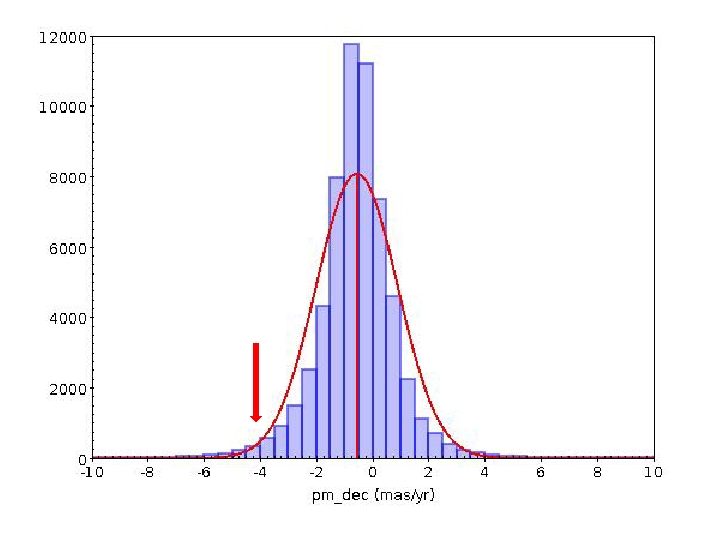}
\caption{ Histogram of the distribution in $\mu_{\delta}$
    (in mas yr$^{-1}$) of the stars within 1 degree of the geometrical center of
    Tycho’s SNR in the range of distance compatible with SN 1572
    (1.7 $<$ d $<$ 3.7 kpc). The data are obtained from {\it Gaia DR2}.
    The red vertical line shows the $\mu_{\delta}$ of star G.}
\label{fig:10}
\end{figure}

\bigskip

\begin{figure}[h!]
\centering
\includegraphics[width=0.6\columnwidth]{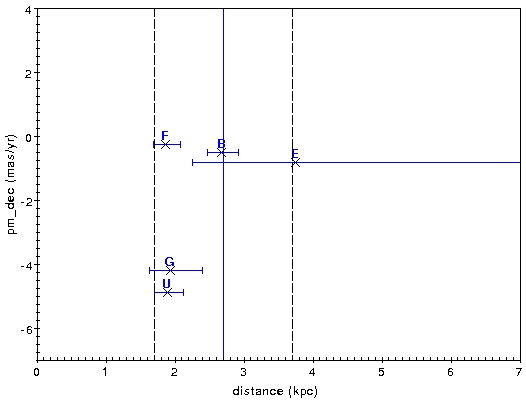}
\caption{ Distances and distance ranges inferred from the parallaxes in the {\it Gaia DR3}   and their uncertainties, together
  with their proper motions in declination. The dashed vertical lines mark the conservative limits of 2.7 $\pm$ 1
  kpc on the
distance to Tycho’s SNR. Solid (blue) error bars correspond to stars from Table 1 that, within reasonable uncertainties, might be
inside the SNR. See for more stars Ruiz--Lapuente et al. (2019), since results
have not changed significantly.}
\label{fig:11}
\end{figure}

\end{document}